\documentclass[10pt,conference]{IEEEtran}
\usepackage{cite}
\usepackage{amsmath,amssymb,amsfonts}
\usepackage{algorithmic}
\usepackage{graphicx}
\usepackage{textcomp}
\usepackage{xcolor}
\usepackage[hyphens]{url}
\usepackage{fancyhdr}
\usepackage{hyperref}
\usepackage{algorithm}
\usepackage{xcolor}
\usepackage{booktabs} 
\usepackage{soul}
\graphicspath{{figs/}}
\usepackage{braket}
\usepackage{etoolbox}

\newcommand{\hpcayear}{2024}

\title{DyPP: Dynamic Parameter Prediction to Accelerate Convergence of Variational Quantum Algorithms}

\def\hpcacameraready{} 

\author{
  \ifdefined\hpcacameraready
    \IEEEauthorblockN{Satwik Kundu, Debarshi Kundu, and Swaroop Ghosh}
    \IEEEauthorblockA{Department of Computer Science and Engineering \\ Pennsylvania State University \\
    \{satwik, dqk5620, szg212\}@psu.edu}
  \else
    \IEEEauthorblockN{\normalsize{HPCA \hpcayear{} Submission
      \textbf{\#\hpcasubmissionnumber{}}} \\
      \IEEEauthorblockA{
        Confidential Draft \\
        Do NOT Distribute!!
      }
    }
  \fi 
}

\fancypagestyle{camerareadyfirstpage}{%
  \fancyhead{}
  
  \fancyfoot[C]{}
}
\begin{document}
\maketitle

\ifdefined\hpcacameraready 
  \thispagestyle{camerareadyfirstpage}
  \pagestyle{empty}
\else
  \thispagestyle{plain}
  \pagestyle{plain}
\fi

\newcommand{\hpcaheight}{0mm}
\ifdefined\eaopen
\renewcommand{\hpcaheight}{12mm}
\fi


\begin{abstract}
The exponential run time of quantum simulators on classical machines and long queue times and high costs of real quantum devices present significant challenges in the efficient optimization of Variational Quantum Algorithms (VQAs) like Variational Quantum Eigensolver (VQE), Quantum Approximate Optimization Algorithm (QAOA) and Quantum Neural Networks (QNNs). To address these limitations, we propose a new approach, DyPP (Dynamic Parameter Prediction), which accelerates the convergence of VQAs by exploiting regular trends in the parameter weights to update parameters. We introduce two techniques for optimal prediction performance namely, Naive Prediction (NaP) and Adaptive Prediction (AdaP). Through extensive experimentation and training of multiple QNN models on various datasets, we demonstrate that DyPP offers a speedup of approximately $2.25\times$ compared to standard training methods, while also providing improved accuracy (up to $2.3\%$ higher) and loss (up to $6.1\%$ lower) with low storage and computational overheads. We also evaluate DyPP's effectiveness in VQE for molecular ground-state energy estimation and in QAOA for graph MaxCut. Our results show that on average, DyPP leads to speedup of up to $3.1\times$ for VQE and $2.91\times$ for QAOA, compared to traditional optimization techniques, while using up to $3.3\times$ lesser shots (i.e., repeated circuit executions). Even under hardware noise, DyPP outperforms existing optimization techniques, delivering upto $3.33\times$ speedup and $2.5\times$ fewer shots, thereby enhancing efficiency of VQAs.
\end{abstract}

\section{Introduction}\label{introduction}

Variational Quantum Algorithms (VQAs) are widely regarded as the most promising application to attain quantum advantage in Noisy Intermediate-Scale Quantum (NISQ) era. One of the key drivers for this belief is that theoretically, VQAs have been found to have the ability to model complex systems that classical algorithms have difficulty simulating \cite{schuld2020circuit}. However, this has been hard to realize in modern-day NISQ-era computers due to many challenges such as, limited number of qubits and significant amounts of quantum noise (e.g., gate error, decoherence error, readout error, crosstalk error, etc.). On top of that, real quantum devices are expensive, and slow (especially due to long queue depths), and access to them is quite limited obstructing the training of VQA circuits.

VQAs are a class of quantum algorithms that employ a hybrid quantum-classical approach to address a wide range of problems such as optimization \cite{delgado2021variational}, determination of the ground state energy of molecules \cite{peruzzo2014variational}, machine learning tasks such as classification \cite{lloyd2013quantum, rebentrost2014quantum}, etc. Thus, VQAs are often considered the quantum counterparts of highly effective machine learning techniques, like neural networks, which have achieved great success in the classical domain \cite{cerezo2021variational}.
To provide an approximate solution to a problem, most current VQAs employ a variational ansatz, also known as a Parameterized Quantum Circuit (PQC) which is made up of trainable parameters that are generally tuned over time using an optimizer to achieve the desired output. Quantum Neural Networks (QNN) can also be thought of as VQAs since they use a variational ansatz to solve optimization/classification problems. QNNs use PQC as the primary tunable block, the parameters of which are updated to perform the prediction task. PQC is typically made up of several 1-qubit and 2-qubit gates that are used to entangle the qubits and explore the search space.

\subsection{Motivation}
\noindent \textbf{Cost of solving the problem:} Executing VQAs with larger circuits involving more qubits and parameters on real quantum devices can be rather expensive. This is because large-scale quantum circuits require costly quantum computing resources and longer convergence times. As a result, more circuit executions, or shots, are needed to achieve the optimal solution. Shots refer to the number of repetitions of a quantum circuit measurement to obtain a probability distribution of outcomes. The expense of employing high-quality quantum machines is directly proportional to the number of shots taken e.g., cloud services like Amazon's Braket provide users easy access to real quantum machines, but they charge for both the number of tasks and shots used \cite{AmazonBraketPricing}. Consequently, the cost of larger quantum circuits that require more shots to reach the optimal solution, increases. Acceleration of the optimization process of VQAs can decrease the number of total circuit runs needed, ultimately reducing the overall cost of utilizing high-quality quantum devices for solving a problem \cite{gu2021adaptive, phalak2023shot}.

\noindent \textbf{Time required to solve the problem:} In \cite{wang2022qoc}, the authors provided the first practical demonstration of training a PQC on real quantum hardware devices using the parameter shift rule. However, it took 2-5 days to train a relatively simple 4-qubit circuit on the quantum machine. Therefore, many researchers prefer quantum simulators to test new ideas and quantum algorithms (with a small number of qubits) as the simulators can easily run on classical computers. Nevertheless, even if a QNN is trained on a simulator, the training time would still be significantly higher compared to a Deep Neural Network (DNN) with 100 times more parameters than the QNN. This is primarily due to the fact that quantum simulators require the simulation of the quantum state, which grows exponentially with the number of qubits. 
As a result, there is a need to speed up the training of large QML models (on real quantum devices or quantum simulators) for
practically relevant applications like drug discovery \cite{li2021drug}. 

Furthermore, parameter optimization for corresponding variational ansatz or PQCs is challenging in VQAs due to the large number of measurements required for the accurate estimation of observable mean values. This high sampling rate can cause a bottleneck in the algorithm runtime. Therefore, it is necessary to minimize the number of measurements or function evaluations needed to improve the efficiency of parameter optimization for PQCs. In most VQAs, quantum circuits are used in combination with classical computation to solve problems to possibly demonstrate quantum advantage. Since these algorithms generally involve performing multiple iterations of quantum circuit parameter optimization, it can hinder the overall performance of the algorithm \cite{bharti2022noisy}. As a result, accelerating the parameter optimization process is crucial for achieving better performance in these hybrid algorithms.

 Motivated by the need for efficient optimization, we ask: is it possible to devise a method to update parameters over certain training epochs without resorting to the computation of forward pass and gradients? Achieving this could significantly alleviate the computational burden, especially in quantum computing, where even basic computations such as gradient calculation necessitate the execution of quantum circuits. These computations are not only resource-intensive but are also often hindered by extended queue times, slowing down the overall process. To circumvent these computational bottlenecks while simultaneously accelerating the optimization process, we put forth a novel methodology paving the way for more efficient quantum computing paradigms.

\begin{figure}[t]
    \vspace{0mm}
    \centering
    \includegraphics[width=\linewidth]{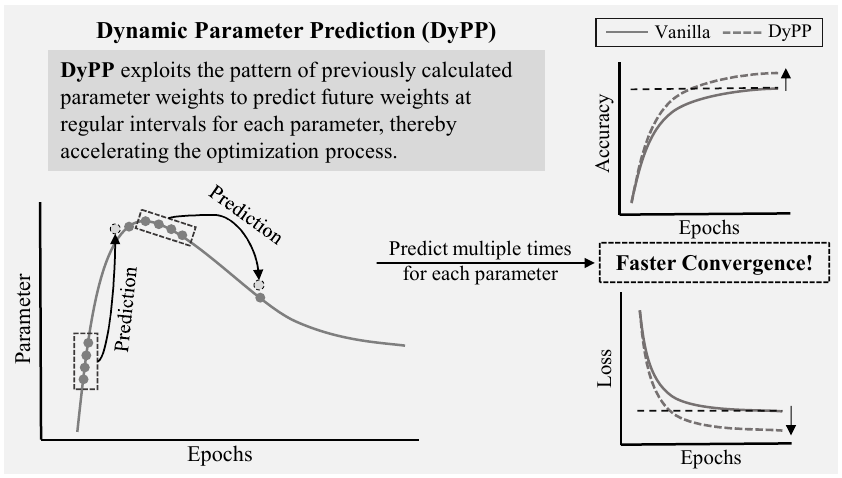}
    \vspace{-2mm}
    \caption{Optimizers update parameters to follow a regular trend as depicted in the plot. DyPP uses this data to predict future weight based on prior trends. Error in predictions are recovered by intermediate optimization using generic optimizers which allows DyPP to make multiple predictions at regular intervals during a training process leading to faster convergence of VQAs.}
    \label{idea}
    \vspace{-4mm}
\end{figure}

\subsection{Proposed Approach}
\textbf{Basic idea:} We present Dynamic Parameter Prediction (DyPP, pronounced as `dip') to accelerate the convergence of VQAs. DyPP involves fitting a non-linear model with a set of previously calculated parameter weights and using it to predict the weight of the parameters for future epochs, which are then updated at regular intervals (Fig. \ref{idea}). This approach enables DyPP to reach the optimal parameters much faster than traditional methodologies, with minimal overhead. 

\textbf{Relevance to quantum computing:} DyPP is particularly well-suited for VQAs compared to classical models such as DNNs. This is because, practically useful DNNs often have billions, if not trillions \cite{openai2023gpt} of parameters, and storing and predicting parameters at regular intervals might greatly increase both the space and time complexity, potentially eliminating the possibility of actual speedup. In contrast, QNNs are known to have superior entanglement capability and effective dimensions \cite{schuld2021effect, abbas2021power, schuld2020circuit} compared to DNNs, requiring only a fraction of parameters e.g., 400 instead of 20k \cite{li2021quantum} to solve any required problem. Entanglement enables quantum models to manipulate data in more complex ways, and a higher effective dimension allows quantum models to represent more complex data. Thus, QNNs, even large ones that are expected for practically relevant applications employing future large-scale quantum hardware, can effectively learn and generalize from smaller amounts of data with significantly fewer trainable parameters than DNNs \cite{schuld2021effect, caro2022generalization}. 

Furthermore, for PQCs, gradient computations methods like the parameter shift rule are costly, necessitating $2 \cdot n$ circuit executions per step for all parameter's gradient, where $n$ is the total parameter count. DyPP circumvents gradient needs, updating parameters using prior weights. This leads to a substantial decrease in total circuit executions, enhancing DyPP's cost-effectiveness and quantum resource efficiency.

 DyPP's curve-fitting approach can potentially mitigate the effects of noise in the calculated parameter weights from prior steps. This can be intuitively understood, as DyPP shares similarities with surrogate-model based optimization (a known noise-resilient technique for optimizing NISQ algorithms \cite{bharti2022noisy, lavrijsen2020classical}) where a surrogate model is constructed using past function evaluations. Essentially, this method 'learns' from previous evaluations to extrapolate and construct an understanding of the overall cost landscape.
In a comparable way, DyPP uses a curve-fitting approach to update parameters based on prior calculations. By fitting a curve to the previously calculated weights, DyPP effectively smoothes out errors in those calculations which can lead to a more reliable optimization.

\textit{To our knowledge, this is the first work that introduces parameter prediction to accelerate the convergence of VQAs.}

In the remainder of the paper, Section \ref{background} provides an introduction to QNN, VQE, and QAOA; Section \ref{methodology} delves into DyPP methodology; Section \ref{evaluation} outlines the evaluation strategies used and discusses the results; Section \ref{discussion} explores potential use cases and addresses the limitations of DyPP; and finally, Section \ref{conclusion} summarizes the key findings.

\begin{figure*}[!t]
    \vspace{-2mm}
    \centering
    \includegraphics[width=0.95\textwidth]{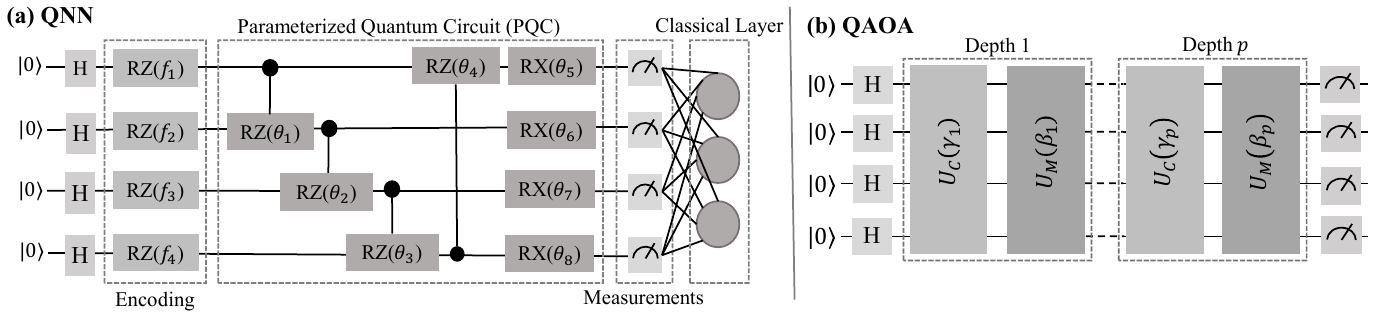}
    \vspace{-3mm}
    \caption{(a) Schematic of a 4-qubit hybrid QNN architecture. Classical features ($f_1, .., f_4$) are encoded as angles of quantum rotation gates ($RZ$ in this case). The PQC transforms encoded states to explore the search space and entangle features. The resulting expectation values are then fed into a classical linear layer for final prediction. (b) A generic QAOA parameterized circuit in which $U_C$ represents the cost layer/unitary with parameters $\gamma_i$ and $U_M$ represents the mixer layer/unitary with parameters $\beta_i$, where $i$ represents the depth of the circuit. Higher depth ($p$) circuit results in a more accurate approximation.}
    \label{qaoa_hqnn_combined}
    \vspace{-4mm}
\end{figure*}

\section{Background} \label{background}

\subsection{Basics of Quantum Computing} 
A quantum bit or qubit is a fundamental building block of quantum computers.
Unlike a classical bit, a qubit can be in a superposition state, which is a combination of states $\ket{0}$ and $\ket{1}$ at the same time. Mathematically, a qubit state is represented by a two-dimensional column vector $[\alpha \; \beta]$ where $|\alpha|^2$ and $|\beta|^2$ represent probabilities that the qubit is in state `0' and `1', respectively. 
Quantum gates are operations that change the state of qubits, allowing them to perform computations. Mathematically, they can be represented using unitary matrices 
There are mainly two types of quantum gates: 1-qubit (like H, X gates) and 2-qubit (like CNOT, CRY gates). Complex 3-qubit gates like Toffoli gates are eventually broken down into 1-qubit and 2-qubit gates during the compilation process.
Qubits are measured on a desired basis to determine the final state of a quantum program. Measurements in physical quantum computers are typically restricted to a computational basis, such as the Z-basis in IBM quantum computers. 
Due to the high error rate of quantum computers, obtaining an accurate output after measuring just once is unlikely. Consequently, quantum circuits are measured multiple times ($n$), and the most probable outcomes from these measurements are then considered as the final output(s). This measurement frequency $n$ is known as \textit{shots}.

\subsection{Quantum Neural Network (QNN)} 
QNN consists of three building blocks: (i) a classical to quantum data encoding (or embedding) circuit, (ii) a PQC whose parameters can be tuned (mostly by an optimizer) to perform the desired task, and (iii) measurement operations. There are a number of different encoding techniques available but for continuous variables, the most widely used encoding scheme is angle encoding where a variable input classical feature is encoded as a rotation of a qubit along the desired axis \cite{abbas2021power}. 
In this study, we used $RZ$ gates to encode classical features into their quantum states.

In QNN, the PQC is the primary trainable block to recognize patterns in data. The PQC is composed of entangling operations and parameterized single-qubit rotations. The entanglement operations are a set of multi-qubit operations (may or may not be parameterized) performed among qubits to generate correlated states and the parametric single-qubit operations are used to search the solution space. 
Finally, the measurement operation causes the qubit state to collapse to either `0' or `1'. We used the expectation value of Pauli-Z to determine the average state of the qubits. The measured values are then fed into a classical layer (number of neurons = number of classes) in our hybrid QNN model as shown in Fig. \ref{qaoa_hqnn_combined}(a), which performs the final classification task.

\subsection{Variational Quantum Eigensolver (VQE)}
VQE \cite{peruzzo2014variational} is a prominent hybrid algorithm primarily used to estimate the ground state energy of a quantum system, represented by a Hamiltonian. 
A Hamiltonian is an observable (depicted by a matrix) associated with the total energy of a system. Its eigenvalues correspond to the different possible energies of the system, with the lowest eigenvalue representing the system's ground state energy. 
It involves following steps; (1) Find the molecular Hamiltonian using the molecules symbol and the atomic coordinates, (2) Choose a PQC with tunable parameters that approximate the true ground state of the molecule. This PQC is used to produce the trial state, with which the molecular Hamiltonian can be measured, (3) Use a classical optimizer to update the parameters of PQC minimize the expectation value of the Hamiltonian until convergence. Once the optimization converges, the final expectation value of the molecular Hamiltonian represents the approximate ground state energy of the molecule. The objective of the VQE is therefore to find a state $\ket{\psi}$, such that the expectation value of the Hamiltonian is minimized \cite{tilly2022variational}. 

\begin{figure*}[!t]
    \vspace{-4mm}
    \centering
    \includegraphics[width=0.9\textwidth]{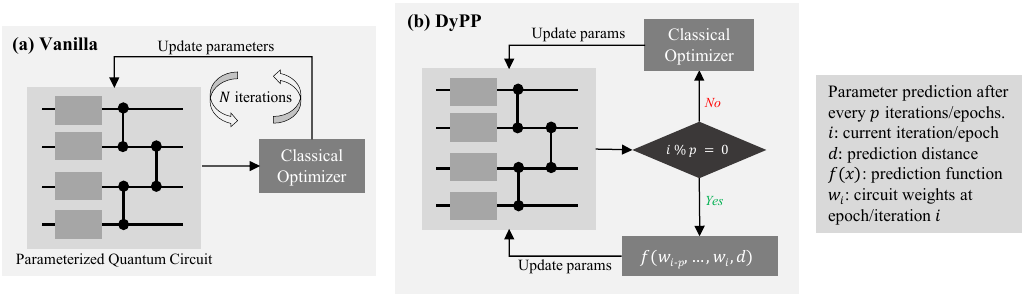}
    \vspace{-2mm}
    \caption{VQAs involve encoding the necessary problem as a cost function into a PQC and evolving the state over time using a set of tunable quantum gates in order to minimize or maximize the expected output. (a) The generalized approach is to use a classical optimizer to update the parameters of the PQC for $x$ iterations to get the required result. (b) DyPP updates the parameters using a prediction function after fitting it with previously calculated weights for each parameter after every $p$ epoch (i.e., when $i \% p = 0$) to accelerate the optimization process.}
    \label{pipeline}
    \vspace{-4mm}
\end{figure*}

Several PQC (ansatz) architectures are proposed for the VQE process such as Hardware Efficient Ansatz (HEA) \cite{kandala2017hardware}, Hamiltonian Variational Ansatz \cite{wecker2015progress} and Efficient Symmetry Preserving (EPS) ansatz \cite{gard2020efficient}. In this work, we used the Unitary Coupled-Cluster Single and Double (UCCSD) excitation variational ansatz \cite{peruzzo2014variational, romero2018strategies} which is described as a superposition of all possible single and double excitations from a reference state (Hartree-Fock \cite{echenique2007mathematical} state), where a single excitation means that one electron is excited from one spin-orbital to another, and a double excitation means that two electrons are excited simultaneously. Intuitively, these excitation gates take lower energy electrons to be in superposition with higher energy electrons and vice versa. The single and double excitation gates are 2-qubit and 4-qubit gates respectively. 

\subsection{Quantum Approximate Optimization Algorithm (QAOA)}
QAOA is a promising hybrid algorithm for finding approximate solutions to combinatorial optimization problems \cite{farhi2014quantum, zhou2020quantum}. In general, any combinatorial optimization problem can be formulated as a cost function $C$($z$) defined using $n$ bit strings $z = (z_1, ..., z_n) \in \{\pm 1\}^{n} $ where the goal is to find $z^{*}$ which results in maximum $C(z^{*})$. The main advantage of QAOA is that it 
scales linearly with the number of qubits used in the circuit. Therefore, it can potentially solve large-scale optimization problems unlike classical optimization algorithms that suffer from exponential scaling. 

A QAOA circuit as shown in Fig. \ref{qaoa_hqnn_combined}(b), is made up of two layers/unitaries called the cost and mixer layers. The cost layer, parameterized by $\gamma$, is a unitary operator ($U_C(\gamma_i) = e^{-i\,\gamma_{i}\,H_{C}}$ where $H_C$ refers to the cost Hamiltonian) that encodes the optimization problem into a quantum circuit. Typically, the cost layer is a sum of Pauli-Z operators acting on qubit pairs. This layer promotes the lowest-cost solutions in the optimization problem. The mixer layer, parameterized by $\beta$, is a unitary operator ($U_M(\beta_i) = e^{-i\,\beta_{i}\,H_{M}}$ where $H_M$ refers to the mixer Hamiltonian) that is designed to mix quantum states such that various solutions can be explored. Typically, the mixer layer is a sum of Pauli-X operators acting on each qubit. 
This mixer layer has been chosen because it preserves the computational basis states, allowing the circuit's output to be easily measured in the computational basis. 
Thus, if the depth of the circuit is $p$ then the total number of tunable parameters is $2p$ i.e. $\{\gamma_1, ..., \gamma_p\}$ and $\{\beta_1, ..., \beta_p \}$. 
Even though increasing circuit depth leads to better performance, it also increases depth and gate count in the circuit and which makes it more susceptible to noise \cite{ayanzadeh2023frozenqubits, tannu2022hammer}. Thus, finding the optimal number of layers is a tradeoff between circuit depth and circuit complexity. QAOA can solve several optimization problems \cite{pagano2020quantum, lloyd2018quantum}. 
However,we used QAOA to solve the graph MaxCut problem in this work. 

\section{Proposed Methodology} \label{methodology}
\subsection{Overview}

\begin{figure}[!b]
        \vspace{-6mm}
        \centering 
        \includegraphics[width=0.75\linewidth]{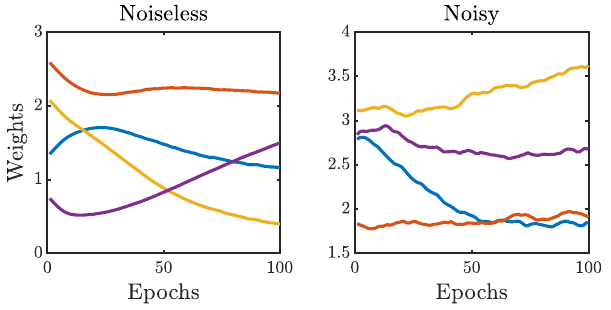}
        \vspace{-4mm}
        \caption{Weight evolution of different quantum parameters of a PQC when training a QNN under both noiseless and noisy environment.}
        \label{weights}
        \vspace{0mm}
\end{figure}

Fig. \ref{pipeline} provides a high-level idea of DyPP. The general idea behind most of the current variational quantum algorithms is to maximize/minimize some cost function value to get the desired output. The practical implementation entails the following steps: design a quantum circuit that encodes the problem the algorithm is trying to solve with a set of tunable parameters and use a classical optimizer \cite{kingma2014adam, ruder2016overview} to update the circuit's weights/parameters to minimize the cost. Thus, the optimizer gradually modifies (i.e., decreases or increases) each of the parameters based on the initialization and cost function at an instance to achieve the required output. Fig. \ref{weights} shows the evolution of parameter weights as a QNN model is trained. It can be noted that even under noise, each parameter exhibits a consistent trend at each instance. We exploit this regular trend in parameters to predict the future parameter weights to speed up quantum algorithms convergence.

To make any meaningful prediction, we need to learn the weight patterns. To do that, in QNN, after training the model for a few epochs, let's say $p$ epochs, we consider weights $\{w_1, .., w_p\}$ ($w_i$ is the parameter weight at epoch $i$) for each of the parameters and use a subset of these weights to fit a curve $f(x)$ (i.e. regression). We use this fitted function to predict a future weight, $w_d = f(d)$ of the parameters ($d$ is the prediction distance, more in Section \ref{pred_distance}). We repeat this prediction process for all the parameters at regular intervals (epochs/iterations) during the training process to accelerate the QNN's training. We define $p$ as the prediction interval i.e., the number of epochs after which we make predictions.

\subsection{Prediction Function}

For predictions, higher-order polynomials are generally effective at approximating complex relationships over a given range of data. Exponential functions require a larger sample size to make accurate predictions whereas linear functions are unsuitable since they are incapable of mimicking a curve such as in Fig. \ref{weights}. Since we are predicting at frequent intervals i.e., after every $p = 4-6$ epochs/iterations, using a high-order polynomial would lead it to overfit making the predictions highly inaccurate. Also, theoretically, given a suitable learning rate, it is unlikely that the curve formed by these small datapoints will have a zero slope at more than one point, making a quadratic polynomial an ideal choice. Still, we  adopted an empirical approach and conducted experiments using linear, cubic, logarithmic, and exponential functions as well. They consistently performed worse or just slightly better than vanilla method (1.05$\times$ speedup in VQE) while the quadratic polynomial significantly outperformed the baseline technique (1.69$\times$ speedup in VQE). Thus, we note that a quadratic function works best, allowing us to make fairly accurate predictions even with a small sample size $p$. As a result, we fit a function of the form $f(x) = ax^2 + bx + c$, 
using weights $[w_2, w_3,.. w_p]$ to get the approximate function coefficients ($a, b, c$) and predict a future weight ($w_d$) using our fitted quadratic function, i.e., $w_d = f(d) = ad^2 + bd + c$, for each parameter. We chose simpler models for weight prediction to minimize space and time complexity, aiming for an optimal balance between the complexity and accuracy of our method. Note that we do not consider the weight $w_1$ and use the remaining $p-1$ weights to fit the curve (discussed in Section \ref{pred_distance}). We define $d$ as the prediction distance.

\subsection{Prediction Distance} \label{pred_distance}
The optimal choice of prediction distance ($d$) so that the predicted weight ($w_d$) is close to the actual future weight is quite difficult to find. Inaccurate predictions can lead to large errors. Also, patterns and trends in weight used for prediction might not hold outside our sample range/size ($p$). Therefore, the value of $d$ should be decided carefully. 

The value of $d$ is dependent on several factors; (i) the sample size or the number of data points (i.e. prediction interval $p$) used to fit the function plays a role in determining the optimal value for $d$. A larger sample size typically allows the function to learn better and make more accurate predictions with a higher value of $d$, although this may not always hold true in practice, (ii) the slope of the fitted curve ($f^{'}(x)$) also affects the choice of $d$. A steeper slope would indicate that the weights are changing rapidly, allowing for larger predictions, while a flatter slope would suggest that the weight is either converging or exhibiting a curvature, and smaller predictions would be more appropriate, (iii) the derivative of the slope, $f^{''}(x)$, is another factor that impacts the value of $d$. A low value of $f^{''}(x)$ would indicate that the slope is changing slowly, allowing for larger predictions, while a high value of $f^{''}(x)$ would suggest that the slope is varying rapidly, and smaller predictions would be more appropriate, (iv) the learning rate also plays a role in determining the optimal value for $d$. A higher learning rate implies taking larger steps towards the optimal value for each weight. In this scenario, it is necessary to use a smaller prediction distance $d$. If the value of $d$ is too large, the model might overshoot the optimal weight, leading to a high loss value, and (v) the point of time for making the prediction. Early on, we can afford to make larger predictions but later on in the training when the weights of the parameters are closer to the minima, we would prefer to make predictions with lower values of $d$ to avoid surpassing the minima.

Thus, several factors  need to be considered to determine the prediction distance for each parameter at each prediction instance. Here, we formulate two prediction methodologies for DyPP; (a) Naive Prediction (NaP) which is inspired by the concept of learning rate decay where we decay the value of $d$ at each prediction interval, and (b) Adaptive Prediction (AdaP) which is a more robust prediction methodology where we determine the prediction distance $d$ for each parameter prediction, depending on the slope ($f^{'}(x)$), the derivative of the slope ($f^{''}(x)$) and the initial learning rate ($\alpha$).

\subsection{Naive Prediction (NaP)} \label{nap}
Here, we formulate $d$ such that it has a reasonably large value initially but decreases as the optimizer gradually updates the parameters and moves toward the convergence/minima. A similar idea is followed for learning rates where one of the ways to increase the probability of convergence is by implementing learning rate decay where the learning rate is slowly decayed based on some predefined functions (step, exponential etc) or adaptively, depending on the gradient. In NaP, we proposed to dynamically reduce the prediction distance $d$ after every $p$ iteration/epoch as:
\begin{align} \label{eq:1}
    d = r^{(i / p)} \cdot d_0 + (p - 1)
\end{align}
where $r$ is known as the decay rate and set to $r = 0.95$,  $i$ is the current epoch/iteration, $p$ is the prediction interval, and $d_0$ is the initial prediction distance.
In the above formula for $d$, we predict after every $p$ epochs (and not $p-1$) even though we use $p-1$ samples to make the predictions. This is because, if the weights obtained just after prediction are slightly off the optimal weight, it may skew the curve fitting process resulting in erroneous predictions. Thus, we train the model for an extra epoch to rectify any inaccuracies caused due to prediction, if any, and use the next $p-1$ epoch weights to make the future prediction. Furthermore, 
if we keep on decreasing the prediction distance with epochs, there might be a time when $d$ becomes less than $p - 1$, in which case we would end up predicting some data within our sample set of weights $p - 1$. To avoid this, we added $p-1$ which is a safe lower bound of the prediction distance $d$. A pseudo-code representation of our proposed DyPP algorithm can be found in Algorithm \ref{algo}.

When working with classical neural network models, we typically need to test models with different hyper-parameters (like learning rate, epochs, etc.) to find the optimal values that will provide the best possible performance. In the above formulation, we define $p$ and $d_0$ as the hyperparameters. The optimal $p$ and $d_0$ would vary depending on the circuit and problem size. However, we provide a safe range of $p$ and $d_0$ as $[(c + 2), 2(c + 2)]$ and $(0, 2p]$, respectively where $c$ is the degree of the polynomial used as the prediction function. The value of $p$ should not be lesser than $c+2$ because, in order to fit a $c$ degree polynomial, we at least need $c + 1$ data points and since we are using $p - 1$ data points for regression, $p - 1 = c + 1$ or, $p = c + 2$. A higher $p$ value might not improve accuracy in noiseless settings with quadratic polynomials, but could be beneficial with higher-order polynomials or in noisy conditions. However, a high $d_0$ value will most likely result in an erroneous prediction, especially in NaP because we will be using the same prediction distance $d$ for each parameter. Additionally, NaP exhibits sensitivity to the learning rate, meaning that as the learning rate $\alpha$ increases, the value of $d_0$ should decrease. To overcome these issues and make a more intelligent prediction we proposed AdaP.

\begin{algorithm}[!t]
\caption{DyPP Algorithm}
\label{algo}
\begin{algorithmic}[1] 
\STATE \textbf{Function} $P_x(W, p)$
\STATE $f(x, a, b, c) \leftarrow ax^2 + bx + c$, $x_n \leftarrow [1, 2, ..., p]$ \COMMENT{Input and function}
\STATE $W_t \leftarrow W.T$, $N \leftarrow len(W_t)$ \COMMENT{Transpose and length}
\STATE Initialize $W_p[N]$ \COMMENT{Array for weights}
\FOR {$i = 1$ to $N$}
    \STATE $a, b, c$ = fit$(f, x_n, W_t[i])$
    \STATE Find prediction distance $d$ using NaP or AdaP
    \STATE Compute and store the predicted weight:
    \STATE $W_p[i] = f(d, a, b, c)$    
\ENDFOR
\RETURN $W_p$
\end{algorithmic}

\begin{algorithmic}[1]
\REQUIRE Number of epochs $E$ and prediction interval $p$.
\ENSURE Optimized PQC weights $W$
\STATE Initialize weights $W$, for storing epoch weights $W_e$
\FOR{$i = 1$ to $E$}
    \STATE Store model weights: $W_e[i] = W$
    \IF{$i$ \% $p == 0$}
        \STATE Get weights based on the last $p-1$ epoch weights:
        \STATE Compute $P_x(W_e[i - (p - 1):i], p - 1)$ 
        \STATE $W =  P_x(W_e[i - (p - 1):i], p - 1)$ \COMMENT{Update weights}
    \ELSE
        \STATE Update parameter weights $W$ using an optimizer
    \ENDIF
\ENDFOR
\RETURN $W$
\end{algorithmic}
\end{algorithm}

\subsection{Adaptive Prediction (AdaP)} \label{adap}
In AdaP, unlike NaP, which uses the same value of prediction distance $d$ for all parameters, the value of $d$ is adaptively determined for each parameter. This adaptive approach reduces incorrect predictions by optimizing the prediction distance for each parameter thereby improving accuracy. In Section \ref{pred_distance} we discussed the relationship of $d$ with the slope, the derivative of the slope, and the learning rate. Intuitively, we can understand that the prediction distance $d$ should be directly proportional to the slope $f^{'}(x)$ since a higher value indicates that the weights are rapidly increasing/decreasing, and thus we can afford to make predictions with a high $d$, whereas if $f^{'}(x)$ is low then we would want to make safer predictions (i.e., low $d$) since it would mean we are either near a curvature or reaching the convergence value. $f^{''}(x)$ should be inversely proportional to $d$ since a low $f^{''}(x)$ indicates that the slope is not varying much which would allow us to make predictions with high $d$ and vice versa. Similarly, learning rate $\alpha$ should also be inversely proportional to $d$ (Section \ref{pred_distance}). Note, by $f^{'}(x)$, we refer to the slope of $f(x)$ fitted by $p-1$ prior weights of the parameter and not the slope/gradient with respect to the loss function.
AdaP uses these three factors ($f^{'}, f^{''}, \alpha$) to adaptively determine the prediction distance for each parameter. We formulate the above relationship as:
\begin{align*}
    &&d_0 \propto &\frac{|f^{'}(x)|}{|f^{''}(x)| \cdot \alpha}&& \\
or, &&d_0 = k \cdot &\frac{|f^{'}(x)|}{|f^{''}(x)| \cdot \alpha + \epsilon}&&
\end{align*}
where $k$ is the proportionality constant or hyperparameter in our case and $\epsilon$ is a constant, which is set to $10^{-6}$ and is defined to prevent numerical instability that may arise from calculations involving very small numbers i.e., when $f^{''}(x) \approx 0$. In the above formulation of $d_0$, we take absolute values of $f{'}(x)$ and $f{''}(x)$ since the direction would not have any separate impact on $d_0$ as we aim to determine how far we can accurately predict irrespective of direction. 

\begin{figure*}[t]
        \vspace{0mm}
        \centering 
        \includegraphics[width=0.85\linewidth]{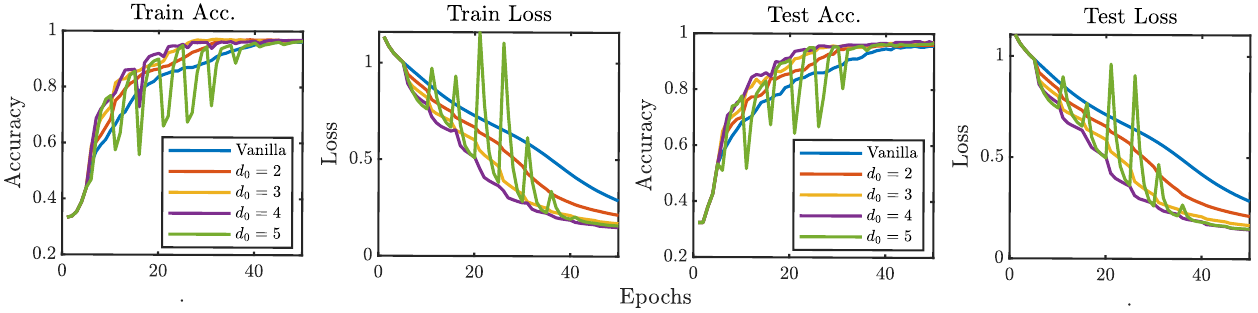}
        \vspace{-3mm}
        \caption{Accuracy and loss of a hybrid QNN model with various starting prediction distance $d_0$ for NaP when training a QNN for 50 epochs on a reduced MNIST 3-class dataset. NaP consistently outperforms original naive training, especially when $d_0$ is in the range [3, 4).}
        \label{d0_variation}
        \vspace{-4mm}
\end{figure*}

However, we cannot use the above formulation of $d_0$ to directly make predictions since the range of $d_0$ would be $[0, \infty)$, making our predictions extremely erroneous for high values of $d_0$. Ideally, we would like to limit the value of prediction distance $d$ to a small range $[0, n)$, so that when $d$ is high (low) it is limited by $n$ (0). There are several ways to map a variable to a finite range. Here, we used an exponential function to map the value of $d_0$ to the range $[0, n)$. Also, as discussed earlier in NaP, we need to set the lower bound of $d$ to avoid predicting within our sample set. Thus, using $d_0$, $n$ and $p$ we formulate our prediction distance $d$ as:
\begin{align*}
    d = (1 - e^{-d_0}) \cdot n + (p - 1)
\end{align*}
where $d_0$ is defined previously, $n$ is the maximum value of prediction distance $d$ and $p$ is the prediction interval.
Now, the only variables which we need to define are the values of $f^{'}(x)$ and $f^{''}(x)$ before we can implement AdaP. We have the quadratic function $f(x) = ax^2 + bx + c$, and for each parameter, after fitting the function using the required weights, we can obtain the coefficients of $f(x)$, i.e. the values of $a$, $b$, and $c$. Therefore, we can calculate $f^{'}(x)$ and $f^{''}(x)$ as follows:
\begin{align*} 
    f^{'}(x) &= \dfrac{d}{dx} (ax^2 + bx + c) = 2ax + b\\
    f^{''}(x) &= \dfrac {d^{2}}{dx^{2}} = 2a
\end{align*}
The value of $f^{''}(x)$ is a constant but the value of the slope i.e. $f^{'}(x)$ is dependent on $x$. Since we predict weight at some future $x$, it makes sense to use the slope at the last datapoint with which we fit the curve. Thus, we take the slope for $f^{'}(p - 1)$ since we fit $f(x)$ with $x = [1, 2, ..., p - 1]$ and $y = [w_2, w_3, ..., w_p]$. Therefore, the final equations used for AdaP methodology are:
\begin{align} \label{eq:2}
    &&d &= (1 - e^{-d_0}) \cdot n + (p - 1)&& \\
    \text{where,} &&d_0 &= k \cdot \frac{|f^{'}(x)|}{|f^{''}(x)| \cdot \alpha + \epsilon}&& \notag \\
    &&f^{'}(x) &= 2a(p-1) + b \text{ and } f^{''}(x) = 2a&& \notag
\end{align}
In the above equation, we define $k$, $p$, and $n$ as the hyperparameters. We found $n = 12$ to work best in most cases. The optimal values of $k$ and $p$ vary depending on the circuit and problem size. However, we provide a safe range for $k$ and $p$ as $(0, 1]$ and $[c + 2, 2(c + 2)]$ respectively where, $c$ refers to the degree of the polynomial we use as our prediction function. A higher $k$ would result in larger predictions (high $d$) more frequently, which can sometimes lead to incorrect predictions. While $p$ can take higher values, it does not necessarily improve prediction accuracy, especially when using quadratic polynomials in noiseless conditions. Higher $p$ values are more beneficial under noisy conditions or with higher-order polynomials, as they can otherwise overfit and produce inaccurate predictions.

\section{Evaluation} \label{evaluation}

\subsection{Setup} \label{setup}
All noiseless simulations are performed using Pennylane's \cite{bergholm2018pennylane} `lightning.qubit" device with the adjoint differentiation method for all gradient calculations (if not mentioned otherwise). For all our prediction/regression tasks, we use the SciPy \cite{virtanen2020scipy} libraries $\mathrm{curve\_fit()}$ function which implements Non-Linear Least Squares (NLLS) optimization using the Levenberg-Marquardt algorithm. Number of shots is set to 1000 for all experiments. All numerical experiments are run on an Intel Core-i7-12700H CPU with 40GB of RAM.

\textbf{Setup for QNN:} For evaluation, we used the most complex circuit (w.r.t number of parameters) i.e., PQC-6 where PQC-x represents circuit-x in \cite{sim2019expressibility}. For embedding classical features to their corresponding quantum state we use the angle encoding technique with $RZ$ gate and for measurement, we calculate the Pauli-Z basis expectation value over all the qubits.

We conduct all experiments using a reduced feature set of MNIST \cite{yan1998mnist}, Fashion \cite{xiao2017fashion}, Kuzushiji \cite{clanuwat2018deep}, and Letters \cite{cohen2017emnist} datasets with latent dimension $d = 8$ (from the original 28$\times$28 sized image) using a convolutional autoencoder \cite{alam2021quantum}. Thus, for each dataset, we create a smaller 4-class dataset from these reduced feature sets i.e., MNIST-4 (class 0, 1, 2, 3), Fashion-4 (class 6, 7, 8, 9), Kuzushiji-4 (class 3, 5, 6, 9) and Letters-4 (class 1, 2, 3, 4) with each having 1000 samples (700 for training and 300 for testing). Since we use 4-qubit QNN models for training and each of these datasets is of dimension $d = 8$, we encode 2 features per qubit. For training, we use the following hyperparameters (unless specified otherwise); epoch = 200, batch\_size = 32, optimizer = Adam \cite{kingma2014adam}, Loss\_fn = SparseCategoricalCrossentropy and learning\_rate ($\alpha$) = 0.002. We use accuracy and loss as metrics to evaluate/analyze the efficacy of our methodology. The hyperparameters for NaP and AdaP methodologies are as follows: $p = 5$, $d_0 = 3$, $k = 0.0001$, and $n = 12$ (Eq. \ref{eq:1}, \ref{eq:2}).

\begin{table*}[t]
    \centering
    \caption{Comparison of final test accuracy and loss after training a hybrid QNN on different datasets for 200 epochs with and without the DyPP (with $p = 5$, $d_0 = 3$ and $k = 0.0001$). AdaP achieves the highest test accuracy and lowest test loss for all the datasets.}
    \label{acc_loss}
    \vspace{-2mm}
    \begin{tabular}{ccccccc}
    \cmidrule(lr){2-7}
    \multicolumn{1}{c}{} & \multicolumn{2}{c}{Vanilla} & \multicolumn{2}{c}{NaP} & \multicolumn{2}{c}{AdaP} \\
    \cmidrule(lr){1-7}
    Datasets    & Test Accuracy & Test Loss & Test Accuracy & Test Loss & Test Accuracy & Test Loss \\
    \cmidrule(lr){1-7}
    MNIST-4     & 0.920      & 0.242      & 0.923     & 0.241     & \textbf{0.943}      & \textbf{0.184}      \\
    Fashion-4     & 0.923      & 0.298      & 0.923      & 0.295     & \textbf{0.947}      & \textbf{0.237}        \\ 
    Kuzushiji-4   & 0.823      & 0.615      & 0.823     & 0.612     & \textbf{0.826}      & \textbf{0.605}       \\ 
    Letters-4   & 0.870      & 0.441      & 0.860     & 0.432     & \textbf{0.876}      & \textbf{0.391}        \\ 
    \bottomrule
    \end{tabular}
\end{table*}

\begin{figure*}[t]
    \vspace{-3mm}
    \centering
    \includegraphics[width=0.9\textwidth]{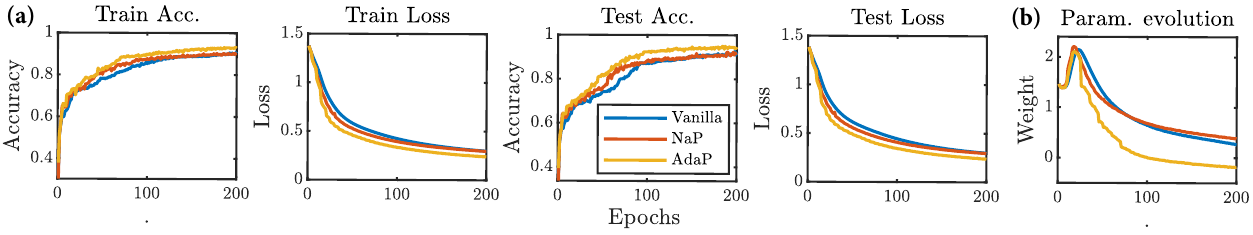}
    \vspace{-3mm}
    \caption{(a) The plot displays a comparison of the performance of our proposed methodologies with vanilla training on the Fashion-4 dataset. The results indicate that our adaptive methodology outperforms all other methods, while the naive prediction methodology also shows slightly better performance than vanilla training. (b)  Weight plot of a parameter in the QNN illustrating how DyPP is able to predict future values early on.}
    \label{fashion_plot}
    \vspace{-5mm}
\end{figure*}

\textbf{Setup for VQE:} In order to find the ground state energy of molecules, we use the UCCSD ansatz. 
We conducted experiments to calculate the ground state energy of various molecules namely, H$_2$, H$_3^{+}$, H$_2$O, BeH$_2$, LiH, NH$_3$, BH$_3$, and CH$_4$ to compare the efficacy of DyPP against vanilla optimization technique. To obtain the molecular Hamiltonian \cite{kim2016pubchem}, we use a subset of electrons and orbitals, enabling quantum simulations with fewer qubits \cite{bao2018automatic}. For experiments, we used the Gradient Descent optimizer with a learning rate ($\alpha$) = 0.1 and convergence tolerance of $10^{-6}$. We initialized all parameters of the circuit to $0$. For the proposed NaP and AdaP methodologies, we set the hyperparameters as follows; $p = 4$, $d_0 = 5$, $k = 0.01$ and $n = 12$ (Eq.\ref{eq:1} and \ref{eq:2}).

\textbf{Setup for QAOA:} For QAOA, we find the approximate MaxCut of different-sized Erdos-Reyni graphs (probability of an edge: 0.6) and compare the performance of DyPP with general technique. We use the approximation ratio as a metric to evaluate the performance of QAOA in solving the MaxCut problem which is defined as: \text{Approximation Ratio (Ar)} = $\frac{\text{ApproxCut}}{\text{MaxCut}}$,
where the `ApproxCut' is the approximate MaxCut predicted by the QAOA circuit and `MaxCut' refers to the actual maximum cut.
Here, we used the Adagrad optimizer \cite{duchi2011adaptive} with a learning rate of 0.05 and ran each experiment for 100 steps. We initialized the parameter values to be uniformly distributed within the range $(0, \pi/2$). The hyperparameters for the proposed NaP and AdaP techniques are set as follows: $p = 4$, $d_0 = 3$, $k = 0.01$, and $n = 12$ (Eq.\ref{eq:1} and \ref{eq:2}).

It should be noted that the goal was not to find the best-performing QNN model \cite{li2023novel}, or the most accurate ground state energy approximation, but rather to demonstrate how DyPP can be easily applied to almost any hybrid algorithms and how well it performs when compared to traditional techniques used in optimizing a quantum circuit. A better choice of hyperparameter or a better prediction distance formulation could in fact provide better performance than shown here.

\begin{table}[!b]
    \vspace{-4mm}
    \centering
    \caption{Speedup provided by DyPP over the vanilla technique for testing accuracy and loss on various datasets.}
    \label{qnn_speedup}
    \vspace{-2mm}
    \begin{tabular}{ccccc}
    \cmidrule(lr){2-5}
    \multicolumn{1}{c}{} & \multicolumn{2}{c}{AdaP} & \multicolumn{2}{c}{NaP} \\
    \cmidrule(lr){1-5}
    Datasets    & Test Acc & Test Loss & Test Acc & Test Loss \\
    \cmidrule(lr){1-5}
    MNIST-4     & \textbf{1.51}      & \textbf{1.48}       & 1.1      & 1.04 \\
    Fashion-4     & \textbf{1.58}       & \textbf{1.55}       & 1.06      & 1.02 \\
    Kuzushiji-4   &\textbf{ 2.25}       & \textbf{1.18}       & 1.30      & 1.04 \\ 
    Letters-4   & \textbf{1.36}       & \textbf{1.74}      & 0.95      & 1.10 \\
    \bottomrule
    \end{tabular}
\end{table}

\subsection{Hyperparameter Analysis} \label{results}
We intuitively determine the prediction interval $p$.
The minimum number of samples needed to fit a quadratic function with 3 variables is 3 i.e., $p = 4$ (since we use the last $p-1$ weights to make the predictions). Therefore, we use $p = 4$ for the VQE and QAOA circuit optimization whereas for QNN training, we note that $p = 5$ can make better predictions.

For NaP, we trained the QNN models with PQC-6 \cite{sim2019expressibility} on a reduced MNIST 3-class dataset (class 0,1,3) with various $d_0 = \{2, 3, 4, 5\}$ for 50 epochs to find the optimal value for $d_0$. To avoid any inconsistencies, we initialized all the models with the same set of weights. Fig. \ref{d0_variation} shows the training and testing loss and accuracy for various values of $d_0$. From the plot, we can note that DyPP consistently outperforms the original/vanilla training. However, not all $d_0$ values are suitable. Low values of $d_0$, e.g., $d_0 = 2$ will result in stable (safe) predictions but will not provide the optimal speed-up. High $d_0$, e.g., $d_0 = 5$ results in several inaccurate predictions (overshooting optimal weight) even though it can regain the accuracy with intermediate training. Thus, an initial prediction distance $d_0$ in the range [3, 4) provides the best results. We use $d_0 = 3$ for the entire QNN training.

For VQE and QAOA, a similar analysis is performed to determine the optimal $d_0$ for NaP. We also conducted experiments with the same QNN model for AdaP to determine the optimal value of hyperparameter $k$ that provides close to optimal performance. As previously discussed in Section \ref{adap}, high values of $k$ result in larger prediction values whereas lower values of $k$ did not provide the optimal performance. As a result, we found that $k = 0.0001$ works best for QNN and $k = 0.01$ gives the best performance in VQE and QAOA. We have also evaluated the effect of different learning rates ($\alpha$) on AdaP's performance for training a QNN (discussed in Section \ref{choices} and shown in Fig. \ref{lr_analysis}).

\begin{figure*}[!t]
    \vspace{-4mm}
    \centering
    \includegraphics[width=0.95\textwidth]{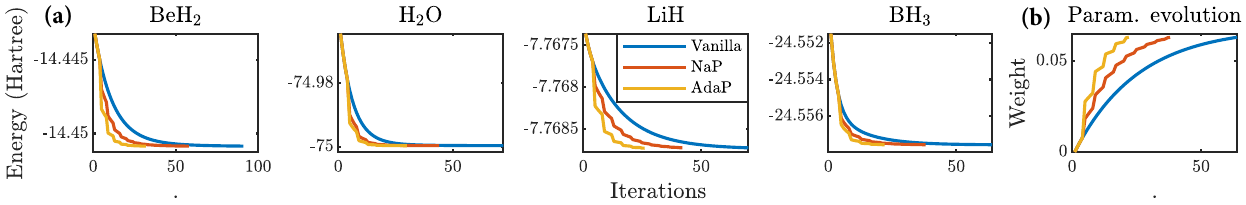}
    \vspace{-3mm}
    \caption{(a) Convergence iterations for different molecules' approximate ground state energies. AdaP excels in all cases with the quickest convergence, succeeded by NaP, and vanilla displays the longest convergence duration. (b)  Weight evolution of a parameter within the VQE circuit for BH$_3$. The plot highlights the effectiveness of DyPP in predicting future parameter values ahead of time.}
    \label{vqe_results}
    \vspace{-5mm}
\end{figure*}

\subsection{Speedup}
We use the following \textit{Speedup} metric to show how quickly DyPP achieves the peak performance compared to the vanilla technique; \textit{Speedup} = $\dfrac{e_v}{e_p}$,
where $e_v$ denotes the epoch/iteration at which the circuit achieves its maximum performance, in terms of loss (accuracy/energy/approximation ratio), using vanilla training, and $e_p$ denotes the epoch at which DyPP surpasses the peak performance achieved through vanilla training.

\textbf{QNN:} For evaluation, we trained QNN with PQC-6 for 200 epochs on multiple datasets. Table \ref{acc_loss} shows the comparison in final testing loss and accuracy between vanilla training, NaP, and AdaP. We chose to represent testing results rather than training loss and accuracy since it represents DyPP's generalization capability better on new data. Even though the final loss/accuracy is not the primary criterion for demonstrating the superiority of our method, it is evident that DyPP consistently outperforms the traditional (vanilla) technique for all datasets. We obtain up to 2.3\% higher final test accuracy and 6.1\% lower final test loss as compared to traditional training methodology. Fig. \ref{fashion_plot}(a) gives a clear representation of the superiority of DyPP over baseline training. Fig. \ref{fashion_plot}(b) shows that DyPP predicts parameters earlier than the baseline method, evidenced by its plot being left-shifted version of the vanilla plot. This indicates AdaP's proficiency in predicting future parameters in advance.

Table \ref{qnn_speedup} shows a speedup of up to 2.25$\times$ and 1.6$\times$ on average by DyPP over vanilla training on various datasets. This means that DyPP can attain the final loss/accuracy obtained through vanilla training in less than half the number of epochs. This significantly reduces the circuit training time and conserves computational and usage costs, especially when training on a real high-cost quantum device as we reduce the total number of circuit executions.

\begin{table}[!b]
    \vspace{-5mm}
    \centering
    \caption{Speedup and the convergence rate provided by NaP and AdaP over the vanilla technique for various molecules.}
    \label{vqe_rate_speedup}
    \vspace{-2mm}
    \begin{tabular}{cccccc}
    \cmidrule(lr){1-6}
    \multicolumn{1}{c}{\textbf{VQE}} & \multicolumn{2}{c}{Speedup} & \multicolumn{3}{c}{Convergence Rate ($\times 10^{-4}$)}  \\
    \cmidrule(lr){1-6}
    Molecules   & AdaP & NaP & Vanilla & AdaP & NaP\\
    \cmidrule(lr){1-6}
    H$_2$       & \textbf{2.25} & 1.60 & 2.38       & \textbf{6.79}     & 4.20  \\
    H$_3^{+}$   & \textbf{2.16} & 1.62 & 2.93       & \textbf{7.64}     & 5.16  \\
    H$_2$O      & \textbf{2.50} & 1.66 & 2.15       & \textbf{7.52}     & 4.04  \\
    BeH$_2$     & \textbf{2.96}  & 1.58 & 0.49      & \textbf{1.78}     & 0.77 \\
    LiH         & \textbf{2.83}  &  1.69 & 0.14      & \textbf{0.45}    & 0.24 \\ 
    NH$_3$      & \textbf{2.12} &  1.69 & 2.03       & \textbf{5.36}     & 3.94  \\
    BH$_3$      &  \textbf{3.10} & 1.72  & 0.39      & \textbf{1.96}     & 0.81  \\
    CH$_4$      & \textbf{2.00}  & 1.71  & 2.78       & \textbf{6.67}     & 5.47  \\
    \bottomrule
    \end{tabular}
\end{table}

\textbf{VQE:} Fig. \ref{vqe_results}(a) shows the iterations required to reach the ground state energy for the benchmark molecules using various methodologies. The reason AdaP requires fewer iterations to converge to the optimal ground state energy can be discerned from Fig. \ref{vqe_results}(b). Table \ref{vqe_rate_speedup} provides the speedup provided by NaP and AdaP for multiple molecules. We note that for BH$_3$, AdaP provides $3.1\times$ speedup whereas NaP provides $1.72\times$ speedup over vanilla technique.
On average, AdaP provides a speedup of $\approx 2.5\times$, and NaP provides a speedup of $\approx 1.65\times$ demonstrating the superiority of DyPP over traditional techniques currently used in the hybrid algorithms.

\begin{figure}[!b]
    \vspace{-2mm}
    \centering
    \includegraphics[width=0.9\linewidth]{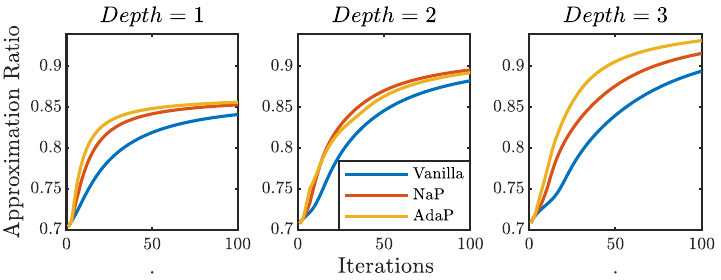}
    \vspace{-3mm}
    \caption{Approximation ratio vs. iteration for an 8-node Erdos-Reyni graph with varying QAOA depths. AdaP and NaP appear to perform better than the traditional optimization approach for depth = 1 and 2, respectively. But, at depth = 3, with more tunable parameters, it is evident that AdaP performs far better than both NaP and the traditional approach.}
    \label{qaoa_results}
    \vspace{-2mm}
\end{figure}

\textbf{QAOA:} Fig. \ref{qaoa_results} shows the change in approximation ratio with time when solving the MaxCut problem for an Erdos-Reyni 8-node graph with different circuit depths. For the lower-depth QAOA circuits (depths 1 and 2) both of the proposed DyPP methodologies equally outperform the vanilla optimization technique. For circuit depth = 3, AdaP performs significantly better than NaP and the traditional technique. Table \ref{qaoa_rate_speedup} shows the speedup provided by AdaP and NaP for different-sized graphs and different circuit depths. AdaP performs the best except for the 8-node depth-2 graph.

\subsection{Convergence Rate (CR)}
We also use \textit{Convergence Rate} or CR as a metric to measure how quickly DyPP converges to optimal solutions and compare them to the vanilla technique. CR, calculated as the absolute value of the loss function's slope over epochs or iterations, is estimated through a linear regression model applied to the loss values. Selecting the appropriate number of epochs is crucial to obtain an accurate CR, as too few epochs may be affected by initial random weights or learning rate adjustments, while too many epochs risk saturation influence. A high CR signifies the algorithm's ability to rapidly and effectively reach an optimal solution, conserving time and resources. Though similar, CR and speedup have distinct meanings: CR gauges training efficiency in epochs, while speedup quantifies the actual time reduction for quantum circuit optimization. 

\begin{table}[!b]
    \vspace{-4mm}
    \centering
    \caption{Speedup and the convergence rate provided by NaP and AdaP over the vanilla technique for different Erdo-Reyni graphs with different QAOA circuit depths.}
    \label{qaoa_rate_speedup}
    \vspace{-2mm}
    \begin{tabular}{cccccc}
    \cmidrule(lr){1-6}
    \multicolumn{1}{c}{\textbf{QAOA}} & \multicolumn{2}{c}{Speedup} & \multicolumn{3}{c}{Convergence Rate ($\times 10^{-3}$)}  \\
    \cmidrule(lr){1-6}
    Nodes-Depth   & AdaP & NaP & Vanilla & AdaP & NaP\\
    \cmidrule(lr){1-6}
    4-1       & \textbf{2.60} & 2.02 & 4.34       & \textbf{5.85}     & 5.67  \\
    4-2   & \textbf{2.25} & 1.98 & 2.93       & \textbf{7.64}     & 5.16  \\
    4-3      & \textbf{1.94} & 1.65 & 2.15       & \textbf{7.52}     & 4.04  \\
    8-1     & \textbf{2.91}  & 2.06 & 3.53      & \textbf{5.44}     & 5.08 \\
    8-2         & 1.34  &  \textbf{1.52} & 3.77      & 4.94    & \textbf{5.58} \\ 
    8-3      & \textbf{2.47} &  1.52 & 2.69       & \textbf{6.60}     & 5.18  \\
    \bottomrule
    \end{tabular}
\end{table}

\begin{figure}[!t]
        \vspace{-3mm}
        \centering 
        \includegraphics[width=0.7\linewidth]{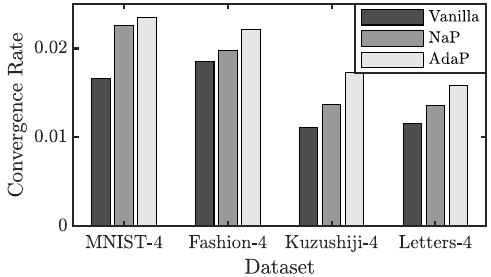}
        \vspace{-3mm}
        \caption{Convergence rate of test loss of QNN for various datasets. DyPP outperforms vanilla training demonstrating accelerated learning.}
        \label{qnn_convrate}
        \vspace{-6mm}
\end{figure}

\textbf{QNN:} Fig. \ref{qnn_convrate} compares the CRs when evaluating QNNs on various datasets. DyPP surpasses the vanilla approach, suggesting that DyPP enables the QNN model to learn underlying patterns within the test data more rapidly and effectively.

\textbf{VQE:} Table \ref{vqe_rate_speedup} compares the CRs of the energy curve of various molecules. Here, we consider energy over all iterations (i.e., iterations required by each method to converge) to calculate the CR. AdaP has up to 400\% higher CR (in the case of BH$_3$) than vanilla methodology. On average, the two DyPP methods outperform the baseline technique significantly i.e. $\approx 223\%$ higher CR (AdaP), and $\approx 83\%$ higher CR (NaP).

\textbf{QAOA:} Table \ref{qaoa_rate_speedup} compares CRs for different graphs and QAOA circuit layers. To calculate CR we consider the approximation ratio corresponding to the first 25 iterations. This is done to avoid the influence of learning rate or saturation on CR. Again, DyPP significantly outperforms vanilla methodology for all cases. AdaP performs the best in all cases except for the 8-node graph with circuit depth of 2, where NaP outperforms both AdaP and vanilla methods. AdaP results in up to $\approx 250\%$ higher CR and NaP provides up to $\approx 88\%$ higher CR compared to the vanilla method.

\begin{table}[!b]
    \vspace{-4mm}
    \centering
    \caption{Table comparing the total number of shots required by various variational quantum algorithms. DyPP requires significantly less number of shots compared to the vanilla optimization technique (cost per shot = $\$0.01$ \cite{AmazonBraketPricing}).}
    \label{shots}
    \vspace{-2mm}
    \begin{tabular}{ccccc}
    \cmidrule(lr){2-5}
    \multicolumn{1}{c}{} & \multicolumn{2}{c}{Shots} & \multicolumn{2}{c}{Savings} \\
    \cmidrule(lr){1-5}
    Algorithms    & Vanilla & DyPP & Shots &  Cost \\
    \cmidrule(lr){1-5}
    QNN     & $2 \times 10^8$       & $\mathbf{1.04 \times 10^8}$ &  1.92$\times$    &  $\$ 960$K\\ 
    VQE     & $5 \times 10^4$       & $\mathbf{1.5 \times 10^4}$ &  3.33$\times$  &   $\$ 0.35$K\\ 
    QAOA    & $1 \times 10^5$       & $\mathbf{0.33 \times 10^5}$ &  3.03$\times$  &   $\$ 0.6$K\\ 
    \bottomrule
    \end{tabular}
    \vspace{-1mm}
\end{table}

\subsection{Shot Reduction}
By using previously calculated weights to update the circuit parameters, we also reduce the total number of circuit executions required to reach the optimal solution. Reduced circuit execution means fewer shots, which leads to significant cost savings. Here, we attempt to provide a lower bound on the number of shots saved per optimization process. In order to approximate the total number of shots, for VQE and QAOA we calculate the total shots as a product of iterations ($N$) required by each on average multiplied by the number of shots for each measurement ($m$) which is set to 1000 by default in Pennylane. For QNN, since we are training with 1000 samples and each forward pass or circuit execution takes in 1 sample, the total number of shots required to train a model is: $\text{shots}_{QNN} = N \cdot \text{samples} \cdot m$

Table \ref{shots} shows that DyPP leads to shot savings of up to $3.33\times$ compared to the vanilla method making DyPP highly (quantum) resource efficient while providing superior performance. Across various VQA tested in this work, DyPP provides an average shot savings of $2.76\times$. This is primarily due to the fact that DyPP avoids function evaluation as it updates the next iterations parameters based on prior weights. As a result, DyPP requires fewer shots overall while providing better performance. The cost of executing circuits on higher quality and/or larger real quantum hardware is directly proportional to the number of shots that a program must execute. As a result, DyPP saves a significant amount of monetary resources, making it extremely cost-effective. 

Note that our shots calculation merely presents a lower bound on the potential savings since it does not account for circuit executions (shots) required for gradient calculations (like in parameter-shift). Considering that DyPP bypasses the need for gradient calculation when predicting (i.e. updating) parameters, its true advantage may be even more pronounced, further enhancing the potential for savings.

\begin{figure}[!t]
        \vspace{-4mm}
        \centering 
        \includegraphics[width=\linewidth]{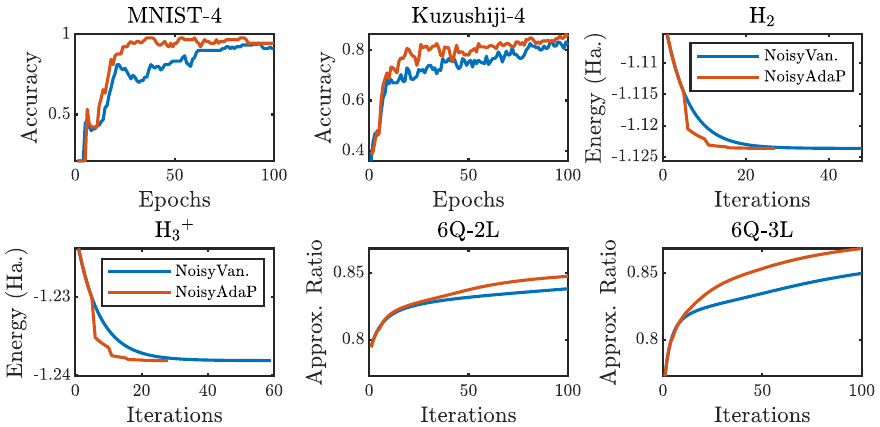}
        \vspace{-6mm}
        \caption{Test accuracy comparison for QNN on MNIST-4 and Kuzushiji-4, VQE ground state energy estimation for H$_2$ and H$_3^{+}$ and QAOA MaxCut performance for a 6-node depth-2/3 graph under noisy conditions. AdaP converges faster than baseline optimization (NoisyVan.) in all the cases.}
        \label{noisy_vqe}
        \vspace{-4mm}
\end{figure}

\subsection{Performance under Noise}
Due to long wait queues of real IBMQ hardware, we used noise-induced simulations on pennylane's ``default.mixed" device for evaluating DyPP's performance under static noise. Since, classical gradient calculation techniques like backpropagation, adjoint, etc. is not feasible on real quantum hardware as intermediate results cannot be stored (due to no-cloning theorem \cite{wootters1982single}), we use parameter-shift (PS) \cite{romero2018strategies, mitarai2018quantum, schuld2019evaluating} and Stochastic Perturbation Simultaneous Approximation (SPSA) \cite{spsa, spsa_pennylane} for gradient estimations. SPSA is a more resource-efficient algorithm since it requires just $2$ circuit execution to approximate the gradient of all parameters, unlike PS which requires $2 \cdot n$ executions ($n:$ num\_params) but it comes at the cost of noisy gradients. Studies have found combining SPSA gradients with other gradient-based optimizers like AMSGrad, Adam, etc. leads to efficient optimization of PQCs in noisy environment \cite{wiedmann2023empirical}. We used SPSA + Adam, PS + SGD, and PS + Adagrad for QNN, VQE and QAOA optimization respectively, keeping default hyperparameters (as in Sec. \ref{setup}) except $p$ which we increased by 1. Fig. \ref{noisy_vqe} compares the performance of vanilla and AdaP in noisy environment. AdaP outperforms vanilla optimization, even under noise, as it reaches optimal accuracy, ground state energy and approx. ratio faster. On average, DyPP provides a speedup of 
2.23$\times$, 1.96$\times$ and 2.1$\times$ for QNN, VQE and QAOA respectively.

To make DyPP more robust to noises, hyperparameters can be adjusted. For example, one can increase the prediction interval $p$ to use more datapoints and fit the model to effectively learn from noisy data and/or limit DyPP's prediction distance ($d$) to only a few steps ahead.
This approach will enable DyPP to perform adequately, but may hinder the speedup. Hardware noise variation can also affect VQAs performance, but DyPP can outperform traditional methods with mitigation techniques \cite{ravi2023navigating, maurya2023scaling, stein2023q} like training with averaged errors \cite{alam2019addressing} or remapping circuits to lowest error subgraphs \cite{nation2023suppressing}. DyPP's aim is not to tackle noise variation, but to accelerate convergence and reduce shots, even under noise conditions.

\section{Discussion} \label{discussion}

\textbf{Usage Model of DyPP:}
DyPP is a versatile plug-and-play tool which can be seamlessly integrated into any VQA to accelerate convergence as long as the VQA uses a PQC whose parameters are being optimized by a classical optimizer and it has access to parameter weights of previous $p-1$ iterations (Algorithm \ref{algo}). To incorporate DyPP into any VQA one can simply check the condition $i \% p == 0$ ($p$: prediction interval) at each iteration/epoch $i$. If false, update weights using the optimizer, else use DyPP (as shown in Fig. \ref{pipeline}). 

The safe ranges of hyperparameters for NaP and AdaP have already been demonstrated for few algorithms and problem sizes. For instance, a smaller value of $k = 0.0001$ is found to be optimal for QNN, whereas a higher value of $k = 0.01$ is more suitable for QAOA and VQE. More generalized ranges are mentioned in Sec. \ref{nap} and \ref{adap}. One could further fine-tune the values of these hyperparameters for general cases using techniques like grid search or random search. 

\begin{table}[!t]
    \vspace{-4mm}
    \centering
    \caption{Performance of AdaP in terms of test loss with increasing QNN width (qubits), while training on 2000 MNIST-10 samples ($\alpha$=0.005). DyPP consistently performs better than vanilla even with increase in quantum circuit size \& complexity.}
    \label{scalability}
    \vspace{-2mm}
    \begin{tabular}{cccc}
    \cmidrule(lr){1-4}
    Width    & Parameters & Speedup & Shots Savings \\
    \cmidrule(lr){1-4}
    4-Q     & 68        & 1.16      & 1.47$\times$  \\
    8-Q     & 168       & 1.78      & 2.22$\times$   \\
    12-Q    & 300       & 1.54      & 1.92$\times$    \\
    \bottomrule
    \end{tabular}
    \vspace{-4mm}
\end{table}

\textbf{Scalability:} 
DyPP's flexibility allows extension to larger QNN models, QAOA implementations with hundreds of qubits, and VQE execution for determining large protein molecules' ground state energy. Table \ref{scalability} showcases the consistently superior performance of DyPP over the vanilla training on varying QNN sizes on a larger MNIST-10 dataset while requiring fewer shots. DyPP exhibits efficient convergence even as circuit size—both in terms of qubits and parameters—increases, and the complexity of the landscape grows because the proposed AdaP strategically determines an optimal prediction distance by considering the slope of the fitted curve and its derivative (or curvature). Thus, for larger PQCs where loss landscapes are likely non-convex, AdaP operates similar to Adam, making smaller predictions corresponding to each parameter's rapidly shifting curvature. 

Yet, DyPP's fundamental idea of fitting a curve for each parameter's prediction promises superior performance with the right hyperparameter tuning. Hence, DyPP should at least equal the baseline's (solely optimizer-based) performance for the same number of shots.
To illustrate this, we examined the worst-case scenario for DyPP, fixing the value of $d$ in the range (0, 1] for both NaP and AdaP from the first iteration and conducted experiments on QNN and VQE. Such a rare scenario may arise when the landscape is very complex allowing DyPP to make very small predictions (low $d$). Table \ref{scalability_complex} reveals that even in this scenario, DyPP matches baseline's performance using fewer shots, highlighting DyPP's advantages.

\begin{figure}[!b]
        \vspace{-4mm}
        \centering 
        \includegraphics[width=0.95\linewidth]{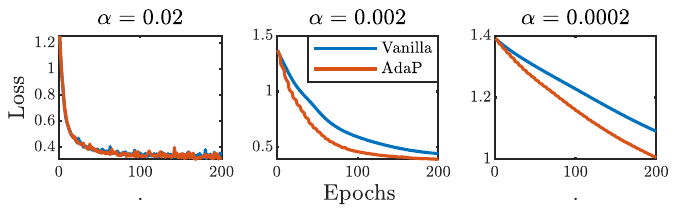}
        \vspace{-2mm}
        \caption{Comparing AdaP vs. vanilla performance (test loss) on Letters-4 with varying learning rates ($\alpha$) in a QNN.}
        \label{lr_analysis}
        \vspace{-1mm}
\end{figure}

\textbf{Impact of Choices:} \label{choices}
DyPP is robust and independent of the specific variational ansatz or the number of layers used in a PQC since it only requires knowledge about prior epochs/iterations weights to update parameters and not the type of gate/parameter. With appropriate hyperparameter tuning, DyPP can perform well for various types of circuits. DyPP can also be augmented with any optimizer to accelerate convergence. In this work, we experimentally demonstrated how well DyPP works when combined with optimizers like, Adam, AdaGrad and SGD using multiple gradient calculation methods (adjoint, spsa, etc.).
The use of a higher learning rate ($\alpha$) may impact the speedup of the DyPP algorithm as can be seen in Fig. \ref{lr_analysis}. However, DyPP still performs comparably to traditional methods in such cases, with its advantage residing in cost and quantum resource efficiency, as it demands fewer total shots for convergence than conventional algorithms.

\begin{table}[!t]
    \vspace{-4mm}
    \centering
    \caption{Performance of DyPP when limiting the prediction distance $d$ in range (0, 1]. Even in the worst case when we are predicting very close by, both AdaP and NaP perform almost similar to vanilla while requiring significantly lesser number of shots.}
    \label{scalability_complex}
    \vspace{-2mm}
    \begin{tabular}{cccc}
    \cmidrule(lr){2-3}
    \multicolumn{1}{c}{} & \multicolumn{2}{c}{Speedup} & \multicolumn{1}{c}{} \\
    \cmidrule(lr){1-4}
    Algorithm    & AdaP & NaP & Shot Reduction\\
    \cmidrule(lr){1-4}
    QNN (4Q)             & 1.02   & 0.95   &  $\sim 1.25 \times$   \\
    QAOA (6Q-3L)        & 0.99     & 0.99   & $\sim 1.33 \times$    \\
    VQE (H$_2$O - 10Q)    & 0.99    & 1.04   &  $\sim 1.31 \times$   \\
    \bottomrule
    \end{tabular}
    \vspace{-4mm}
\end{table}

\textbf{DyPP vs Traditional Optimizers:}
Gradient-based optimizers such as Adam and Adagrad uses gradients to update parameters which is costly especially in quantum domain, while gradient-free optimizers generally require higher number of function evaluations while converging more slowly compared to gradient-based counterparts. In contrast, DyPP leverages prior weights to update parameters. The reliance on previously calculated weights indicates that DyPP cannot independently carry out the optimization process but rather complement an optimizer to achieve high performance efficiently. Any prediction inaccuracies from DyPP are rectified during the intervening optimization steps, underscoring the importance of this balanced and combined approach. Moreover, DyPP's unique design enables it to avoid the necessity of both the next function evaluation and gradient calculation when updating parameters, while potentially accelerating the optimization process. Therefore, using DyPP for intermediate parameter prediction is significantly more efficient than relying solely on generic optimizers, proving particularly beneficial for resource-scarce and time-costly NISQ era. In essence, the power of DyPP becomes evident when it is employed as an enhancement to existing optimization mechanisms. 

Robust optimizers like Adam, AMSGrad adaptively adjust their step size using exponential average of gradient and it's standard deviation. Conversely, DyPP updates weights by fitting a curve $f(x)$ using prior weights, and using its first($f^{'}(x)$) and second($f^{''}(x)$) derivatives to determine prediction distance(AdaP). It should be noted that, $f^{'}(x)$ is fitted curve's slope at the last data point and not the gradient of the parameter w.r.t. loss function. Therefore, DyPP aims to predict parameters based on their trajectory, contrasting with a generic optimizer, which attempts to update based on direction of the gradient (with respect to the loss function).

\textbf{Limitations:} \label{limitations}
The above experiments show that DyPP can accelerate any variational quantum algorithm that uses an optimizer to update the parameters of the required PQC to achieve the desired output. However, DyPP comes at the cost of an increase in space complexity and possibly an increase in time complexity as well depending on the implementation.

\textit{Space Complexity:}
DyPP requires $p-1$ previous weights which would need storage of $N*p$ weights, where N is the total number of parameters in the circuit/model. This would necessitate a higher space complexity of $O(Np)$. However, in most cases a low value of $p$ in the range $[4, 6]$ results in relatively high performance. Therefore, the space complexity would simplify to $O(N)$. 
However, this complexity could impact models with billions of parameters (like in GPT-3 \cite{brown2020language}). Hence, DyPP is more suitable for VQAs where the parameter count is in the thousands. Furthermore, it has been demonstrated \cite{chen2020variational, li2021drug} that QML models can generalize effectively while using a fraction of the parameters required by deep learning models ($\approx 1\%$). Thus, for smaller $N$, the overhead of our methodology is minimal, making DyPP practically viable, particularly in the current NISQ era. One can implement techniques to reduce space complexity, 
however, our focus was to demonstrate updating quantum parameters with prior weights, rather than designing the most efficient implementation of DyPP.

\textit{Time Complexity:} 
Since each prediction instance requires multiple non-linear regression/curve fitting, one might assume that the total time complexity might be really high. However, 
since each of our regressions involves very few data points i.e., $p-1$ to fit a quadratic function, it does not have a drastic impact on the total run time. Furthermore, since each prediction is independent, i.e., the prediction of the future value of the parameter does not require any information about the other parameters, 
we can easily parallelize them across different cores/threads. There are multiple AI parallel libraries like Ray \cite{moritz2018ray}, Horovod \cite{sergeev2018horovod}
etc. which can be used to distribute regression tasks across multiple cores, GPUs, or nodes in a cluster, speeding up the prediction process.

\section{Conclusion} \label{conclusion}
Variational Quantum Algorithms (VQAs) can potentially solve certain problems faster than classical algorithms due to the inherent capabilities of quantum computing but this advantage has not been realized yet. The training time of VQA circuits is currently limited by the computational intensity of quantum simulators and limited resources. In this study, we propose DyPP which leverages regular trends in parameter weights to predict and update the parameter weights of VQAs at regular intervals. The results show that DyPP can significantly speed up the convergence of these hybrid algorithms by up to $3.2\times$ while requiring $2.76\times$ lesser number of shots on average, with only a small additional cost in classical storage and computational resources. 


\bibliographystyle{IEEEtranS}
\bibliography{refs}

\end{document}